\documentclass[aps,prl,reprint,superscriptaddress]{revtex4-2}
\usepackage{graphicx}
\usepackage{graphicx}
\usepackage{dcolumn}
\usepackage{bm}
\usepackage[dvipsnames]{xcolor}
\usepackage{amsmath}

\newcommand{\bfpsi}{\mbox{\boldmath$\psi$}}

\begin{document}

\title{Antiferromagnetic order in Co-doped Fe$_5$GeTe$_2$ probed by resonant magnetic x-ray scattering}

\author{Xiang Chen} 
\email{xiangchen@berbeley.edu}
\affiliation{Materials Science Division, Lawrence Berkeley National Lab, Berkeley, California 94720, USA}
\affiliation{Physics Department, University of California, Berkeley, California 94720, USA}

\author{Enrico Schierle} 
\affiliation{Helmholtz-Zentrum Berlin für Materialen und Energie, BESSY II, D-12489 Berlin, Germany. }

\author{Yu He}
\affiliation{Department of Applied Physics, Yale University, New Haven, Connecticut, 06511, USA}
\affiliation{Physics Department, University of California, Berkeley, California 94720, USA}
\affiliation{Materials Science Division, Lawrence Berkeley National Lab, Berkeley, California 94720, USA}

\author{Mayia Vranas}
\affiliation{Department of Physics, University of California, San Diego, California 92093, USA}

\author{John William Freeland} 
\affiliation{Advanced Photon Source, Argonne National Laboratory, Argonne, Illinois 60439, USA}

\author{Jessica. L. McChesney} 
\affiliation{Advanced Photon Source, Argonne National Laboratory, Argonne, Illinois 60439, USA}

\author{Ramamoorthy Ramesh}
\affiliation{Department of Materials Science and Engineering, University of California, Berkeley, California 94720, USA}
\affiliation{Materials Science Division, Lawrence Berkeley National Lab, Berkeley, California 94720, USA}
\affiliation{Physics Department, University of California, Berkeley, California 94720, USA}

\author{Robert J. Birgeneau}
\affiliation{Physics Department, University of California, Berkeley, California 94720, USA}
\affiliation{Materials Science Division, Lawrence Berkeley National Lab, Berkeley, California 94720, USA}
\affiliation{Department of Materials Science and Engineering, University of California, Berkeley, California 94720, USA}

\author{Alex Frano}
\email{afrano@ucsd.edu }
\affiliation{Department of Physics, University of California, San Diego, California 92093, USA}

\date{\today}

\begin{abstract}

The quasi-two-dimensional van der Waals magnet Fe$_{5-\delta}$GeTe$_2$ has emerged as a promising platform for electronic and spintronic functionalities at room temperature, owing to its large ferromagnetic ordering temperature $T_{\text{C}}$ $\sim$ 315 K. Interestingly, by cobalt (Co) substitution of iron in F5GT, $i.e.$ $({\text{Fe}}_{1-x}{\text{Co}}_x)_{5-\delta}{\text{GeTe}}_2$ (Co-F5GT), not only can its magnetic transition temperature be further enhanced, but the magnetic and structural ground states can also be tuned. Specifically, an antiferromagnetic (AFM) order is induced beyond the Co doping level $x \ge 0.4$. Here, we investigate the magnetic properties of a Co-F5GT single crystal at $x = 0.45(1)$, by utilizing the element specific, resonant magnetic x-ray scattering technique. Our study reveals an A-type, Ising-like AFM ground state, with a transition temperature $T_{\text{N}}$ $\sim$ 340 K. In addition, our work unveils an important contribution from Co magnetic moments to the magnetic order. The application of the in-plane magnetic fields gradually polarize the spin moments along the field direction, but without inducing incommensurate spin texture(s).

\end{abstract}

\maketitle

The unique nature of the cleavable, quasi-two-dimensional (quasi-2D) van der Waals (vdW) materials offers rich platforms for exploring both exotic physical phenomena and technological applications \cite{Park_2016_JPCM, Novoselov_2016_Science, Liu_2016_NatRevMat, Burch_2018_Nature, Gibertini_2019_NatNano, Gong_2019_Science, Mak_2019_NatRevPhy, Du_2021_NatRevPhy, Sierra_2021_NatNano}. A broad variety of intriguing physical phenomena have been reported in recent years by investigating the (atomically thin) vdW bonded compounds. Some exceptional phenomena include unconventional superconductivity in twisted graphene \cite{Cao_2018_Nature}, nonlinear Hall effect in few-layer WTe$_2$ \cite{Ma_2019_Nature, Kang_2019_NatMat}, 2D magnetism in monolayer Cr$_2$Ge$_2$Te$_6$ \cite{Carteaux_1995_JPCM, Gong_2017_Nature}, CrI$_3$ \cite{McGuire_2015_CM, Huang_2017_Nature} and the quantum anomalous Hall effect in insulating tellurides \cite{Chang_2013_Science, Deng_2020_Science}. Among the different materials, the layered vdW, itinerant magnets represent ideal quasi-2D material systems which enable the coupling between the electronic and magnetic degrees of freedom \cite{Park_2016_JPCM, Burch_2018_Nature, Gibertini_2019_NatNano, Gong_2019_Science, Mak_2019_NatRevPhy}. Therefore, vast opportunities arise because of the rich physical properties and the abundant possibilities for functional devices \cite{Huang_2020_NatMat, Guo_2021_JMCC}.

Very recently, some promising quasi-2D magnetic tellurides \cite{Deiseroth_2006_EJIC_FGT, Chen_2013_JPSJ, May_2016_PRM_F3GT, Freitas_2015_JPCM, Sun_2020_NanoRes, Deng_2018_Nature, Li_2018_NanoLett, Seo_2020_SciAdv_F4GT, Stahl_2018_ZFAAC_F5GT, May_2019_ACSnano_F5GT, Zhang_2020_PRB_F5GT, Yamagami_2021_PRB, Wu_2021_PRB}, such as Fe$_5$GeTe$_2$ (F5GT) and CrTe$_2$, with atomically thin nanoflakes for promising room temperature (RT) spintronics have been reported. The F5GT compound has a ferromagnetic (FM) transition above RT at $T_{\text{C}}$ $\sim$ 315 K \cite{May_2019_ACSnano_F5GT, Zhang_2020_PRB_F5GT}. Intriguingly, with cobalt (Co) substitution of iron (Fe) in F5GT ($({\text{Fe}}_{1-x}{\text{Co}}_x)_{5-\delta}{\text{GeTe}}_2$, Co-F5GT), the magnetic transition temperature is further increased up to $\sim$ 360 K, and the magnetic ground state switches from FM to antiferromagnetic (AFM) when the Co doping level reaches $x \ge 0.4$ \cite{May_2020_PRM_FCGT, Tian_2020_APL_FCGT}. Interestingly, a novel wurtzite-type polar magnetic metal, which hosts a metastable, zero-field N$\acute{\text{e}}$el-type skyrmion lattice at RT, was discovered at $x$ = 0.5 of Co-F5GT \cite{Zhang_2021_FCGT, Zhang_2022_FCGT_sciadv}. The aforementioned observations of the Co-F5GT system highlight its immense tunability and capacity for unusual magnetic properties and could render applications in next-generation spintronics.

Despite the reports of exotic magnetic textures such as the chiral soliton lattices and skyrmions in F5GT or Co-F5GT \cite{Zhang_2021_FCGT, Ly_2021_AFM, Gao_2020_AdvMat}, there is a scarcity of information from neutron scattering studies of these systems. The challenge of growing single crystals large enough for neutron experiments hinders the timely investigations of the magnetic properties of Co-F5GT. Here, we demonstrate the applicability of utilizing the element specific, resonant magnetic x-ray scattering (RMXS) technique \cite{Blume1985, Hannon1988PRL, Fink_2013_RPP, Comin_2016_ARCMP} to investigate the bulk magnetic properties of a specific Co-F5GT single crystal at $x$ = 0.45(1). By tuning the x-ray energy $E=\hbar\omega$ to either the Fe $L$ edges or Co $L$ edges, our study determines the contribution from both the Fe and Co spin moments to the magnetic order. In addition, our scattering data affirms the A-type AFM order, with a N$\acute{\text{e}}$el temperature $T_{\text{N}}$ $\sim$ 340 K and an out-of-plane moment direction. The influence of external magnetic fields on the AFM ground state is also explored and found to be consistent with magnetization measurements.

\begin{figure}[t]
\centering
\includegraphics[width= 8.5 cm]{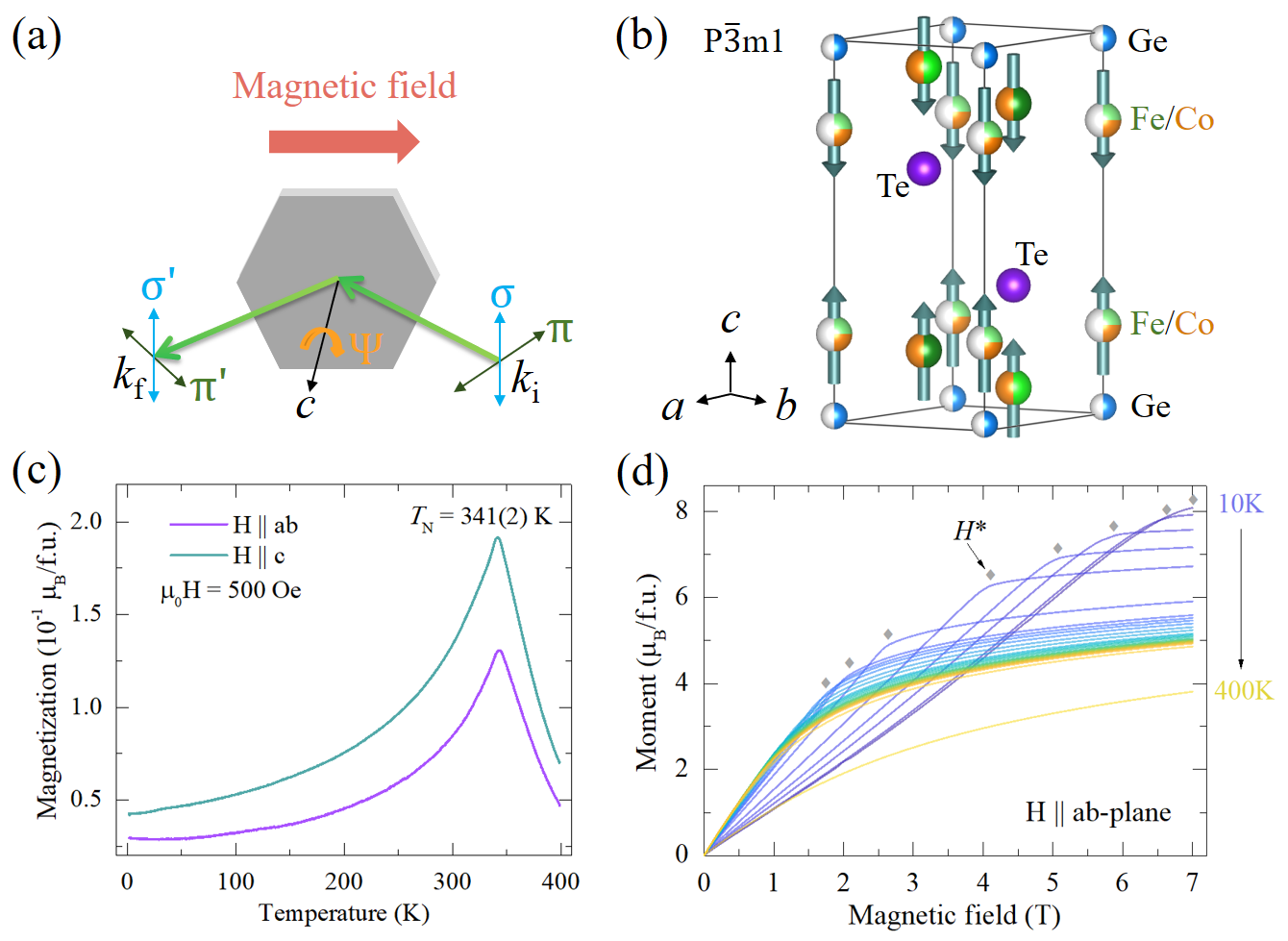}
\caption{(Color online) (a) The schematic of the resonant magnetic x-ray scattering (RMXS) experiment on a Co doped Fe$_5$GeTe$_2$ single crystal at $x = 0.45(1)$ ($({\text{Fe}}_{0.55}{\text{Co}}_{0.45})_{5-\delta}{\text{GeTe}}_2$, Co45-F5GT). (b) The crystal structure of Co45-F5GT, with the AA stacking order \cite{May_2020_PRM_FCGT, Zhang_2021_FCGT, Zhang_2022_FCGT_sciadv}. (c) Temperature dependent magnetization of Co45-F5GT. (d) In-plane isothermal magnetization of Co45-F5GT at varying temperatures. The $H^{\ast}$ is the saturation field, beyond which the spin moments are fully polarized along the field direction with a saturation magnetization $M_{\text{sat}}$. }
\label{fig:Fig1_alpha}
\end{figure}

\begin{figure}[t]
\centering
\includegraphics[width = 5 cm]{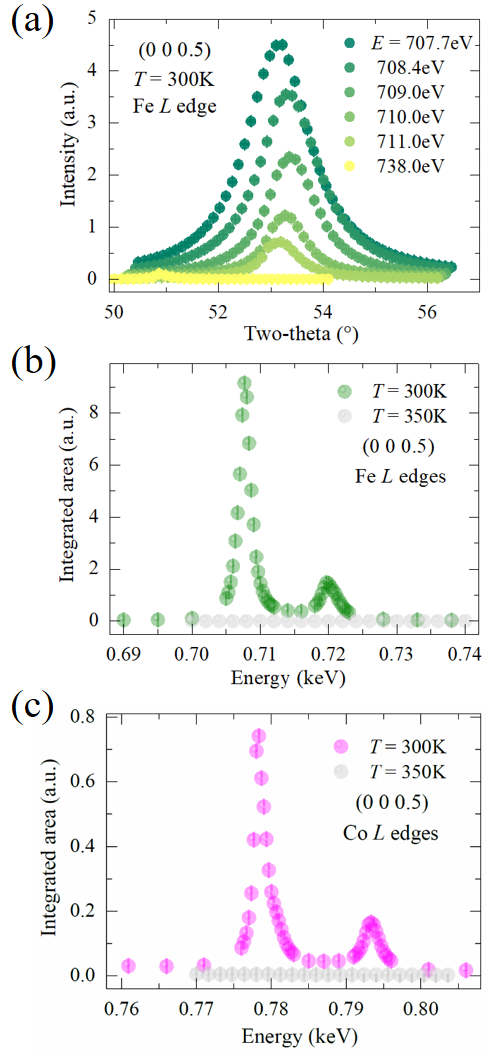}
\caption{(Color online) Photon energy dependence of the (0, 0, 0.5) peak at $T$ = 300 K or 350 K. (a) The (0, 0, 0.5) peak at $T$ = 300 K near the Fe $L$ edges. Maximum intensity is observed at Fe $L_{\text{III}}$ edge ($E$ = 707.7 eV). (b)-(c) Energy dependent integrated area of the (0, 0, 0.5) peak at both 300 K and 350 K, near the Fe $L$ edges (b) and Co $L$ edges (c), respectively.}
\label{fig:Fig2_alpha}
\end{figure}

Single crystals of Co-F5GT at $x = 0.45(1)$ (labeled here as Co45-F5GT) were synthesized using the chemical vapor transfer technique \cite{May_2019_ACSnano_F5GT, Zhang_2020_PRB_F5GT, supp}. The chemical composition of the samples was verified by energy dispersive x-ray spectroscopy, with the cation deficiency $|\delta| \le 0.1$. The RMXS experiments were performed at the UE46$_{-}$PGM-1 beamline at Helmholtz-Zentrum Berlin (HZB) and the 29-ID IEX beamline at Advanced Photon Source (APS) at Argonne National Laboratory. A horizontal scattering geometry is utilized with the sample lattice $c$ direction lying within the scattering plane (Fig. 1(a)). The Bragg peaks $\textbf{Q}$ = $(H\cdot\frac{2\pi}{a}, K\cdot\frac{2\pi}{b}, L\cdot\frac{2\pi}{c})$ are defined in reciprocal lattice units ($r.l.u.$) with lattice parameters $a$ = $b$ $\approx$ 4.02 \AA \, and $c$ $\approx$ 9.80 \AA. The data are collected near the Fe and Co $L$ edges to enhance the magnetic scattering signal \cite{Kao_1990_PRL_Fe_RMXS, Yamasaki_2015_PRB_FeGe_RMXS, Windsor_2017_PRB_CoCr2O4_RMXS}. The incoming x-ray is either horizontally polarized ($H$-pol or $\pi$-pol) or vertically polarized ($V$-pol or $\sigma$-pol), but the scattered x-rays are not analyzed. Therefore, both outgoing $\sigma'$-pol and $\pi'$-pol channels will contribute to the scattered intensity, which can be written as \cite{Fink_2013_RPP, Comin_2016_ARCMP}:

\begin{equation}
I_{\mu \nu} \propto | \sum_{j} e^{ i \textbf{Q} \cdot \textbf{r}_j} ( \textbf{e}_{\mu} \times \textbf{e}_{\nu}^{\ast}) \cdot \textbf{m}_j F(\text{E})|^2,
\label{eq:eqn1}
\end{equation}

to first order in the magnetic moment $\textbf{m}_j$ of the ion located at site $\textbf{r}_j$ within the unit cell. $\textbf{e}_{\mu}$ ($\mu = \sigma$ or $\pi$) and $\textbf{e}_{\nu}$ ($\nu = \sigma'$ or $\pi'$) are unit vectors along the polarization of the electric field component of the incident and out-going x-ray beams, respectively. $F(\text{E})$ is the non-local, photon energy dependent scattering tensor. Since the polarization of the scattered light is not analyzed, the measured intensities are $I_{\text{H}}$ = $I_{\pi}$ $\equiv$ $I_{\pi \sigma'}$ + $I_{\pi \pi'}$ and $I_{\text{V}}$ = $I_{\sigma}$ $\equiv$ $I_{\sigma \sigma'}$ + $I_{\sigma \pi'}$.

The F5GT compound is composed of three identical layers with the rhombohedral layer stacking (space group R$\bar{3}$m), labelled as ABC-stacking \cite{May_2019_ACSnano_F5GT}. It was experimentally established that the stacking order in F5GT is susceptible to external perturbations \cite{May_2020_PRM_FCGT, Zhang_2021_FCGT, Chen_2022_PRL}. Upon replacing Fe with Co in F5GT, when $x \ge 0.4$, the crystal structure undergoes a transition from ABC-stacking to AA-stacking (space group P$\bar{3}$m1, as shown in Fig. 1(b)) \cite{May_2020_PRM_FCGT}. Meanwhile, the magnetic ground state evolves from FM at $x < 0.4$ to AFM when $x \ge 0.4$. Our magnetization measurements on the Co45-F5GT sample confirm an AFM order, with $T_{\text{N}}$ $=$ 341(2) K (Fig. 1(c)). It is worth emphasizing that a unique type of stacking order, labelled as AA$'$-stacking, was reported recently at $x = 0.5$ \cite{Zhang_2021_FCGT, Zhang_2022_FCGT_sciadv}. N$\acute{\text{e}}$el-type skyrmions were identified at this doping, because of the AA$'$-stacking order which breaks the inversion symmetry and therefore features a bulk Dzyaloshinskii–Moriya interaction \cite{Dzyaloshinsky_1958, Moriya_1960}. The spin structures of Co-F5GT, however, have not yet been investigated by bulk scattering techniques. Particularly, the magnetic ground state, including the spin moment direction, shows a strong dependence on the Co doping level $x$ \cite{May_2019_ACSnano_F5GT, Zhang_2020_PRB_F5GT, Zhang_2021_FCGT, Ly_2021_AFM}. Consequently, it is important to investigate the magnetic properties of Co-F5GT by using a direct, bulk-sensitive, element-specific scattering technique such as RMXS.

\begin{figure}[t]
\centering
\includegraphics[width= 5 cm]{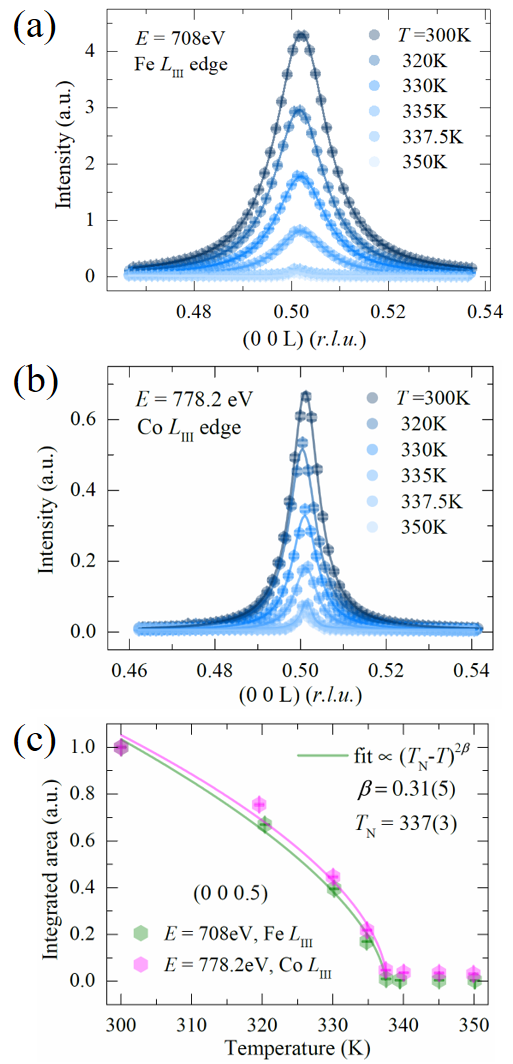}
\caption{(Color online) Temperature dependence of the (0, 0, 0.5) peak, at different energies: (a) Fe $L_{\text{III}}$ edge, $E$ = 708 eV, (b) Co $L_{\text{III}}$ edge, $E$ = 778.2 eV. Solid lines in (a)-(b) are Lorentzian fits to the peak intensity. (c) Power law fits to the integrated area of the (0, 0, 0.5) peak at both Fe and Co $L_{\text{III}}$ edges. For better comparison, the data is normalized to 1 at $T$ = 300 K.}
\label{fig:Fig3_alpha}
\end{figure}

\begin{figure}[t]
\centering
\includegraphics[width= 5 cm]{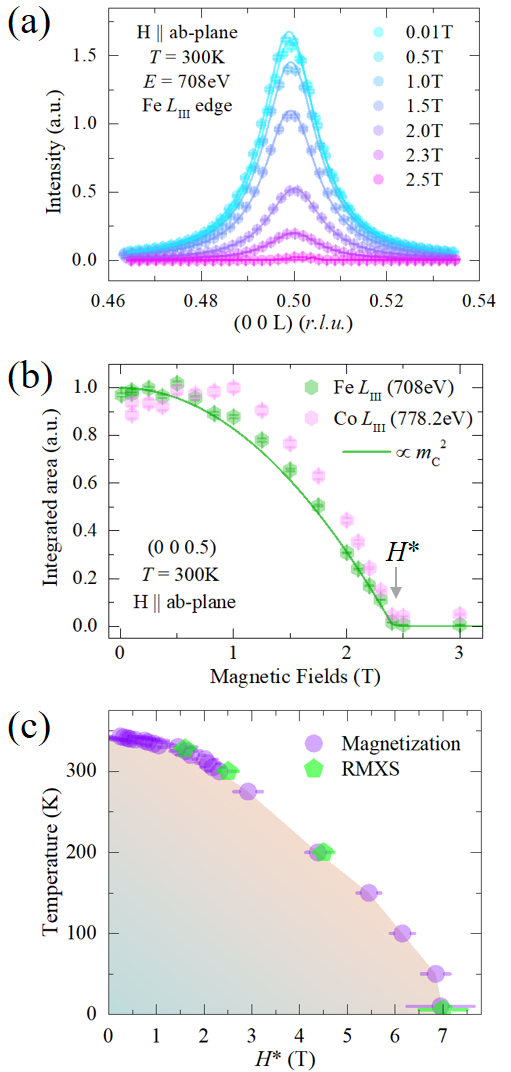}
\caption{(Color online) Magnetic field dependence of the (0, 0, 0.5) peak on resonance. (a) The (0, 0, 0.5) peak at $T$ = 300 K and $E$ = 708 eV, collected under select magnetic fields. Solid lines are Lorentzian fits to the peak intensity. (b) Field dependence of the integrated area of the (0, 0, 0.5) peak at 300 K, at both Fe and Co $L_{\text{III}}$ edges. The peak intensity is completely suppressed beyond $H^{\ast}$ = 2.4(1) T. The green line is a fit to the data by assuming the intensity is proportional to the $c$ component of the magnetization squared $m_{\text{c}}^2$. For better comparison, the data is normalized to 1 at zero field. (c) Phase diagram of the temperature dependent $H^{\ast}$, from both the magnetization and RMXS measurements.}
\label{fig:Fig4_alpha}
\end{figure}

Our RMXS experiments were performed on Co45-F5GT single crystals with the AA-stacking sequence (P$\bar{3}$m1, Fig. 1(b)). A strong peak is identified at the structurally forbidden Bragg peak position $\textbf{Q}_0$ = (0, 0, 0.5) below the magnetic onset temperature. To understand fully the nature of this reflection, we present a thorough study of its dependence on x-ray energy, polarization, temperature and magnetic field. Fig. 2 shows the x-ray energy dependence of the $\textbf{Q}_0$ reflection. Resonant peak profiles were recorded near both the Fe and Co $L_{\text{III}}$ or $L_{\text{II}}$ edges (Fe $L_{\text{III}}$ $\sim 708$ eV, Fe $L_{\text{II}}$ $\sim 720$ eV, Co $L_{\text{III}}$ $\sim 778$ eV and Co $L_{\text{II}}$ $\sim 793$ eV) \cite{Kao_1990_PRL_Fe_RMXS, Yamasaki_2015_PRB_FeGe_RMXS, Windsor_2017_PRB_CoCr2O4_RMXS}. A single energy dependent peak profile is evident near both the Fe and Co $L_{\text{III}}$ ($L_{\text{II}}$) edges. This resonance in photon energy is a strong signature of the magnetic nature of the $\textbf{Q}_0$ peak~ \cite{Fink_2013_RPP, Comin_2016_ARCMP, Kao_1990_PRL_Fe_RMXS, Yamasaki_2015_PRB_FeGe_RMXS, Windsor_2017_PRB_CoCr2O4_RMXS}. To support this argument further, the temperature dependence of the $\textbf{Q}_0$ peak at the resonance energies (both Fe and Co $L_{\text{III}}$ edges) was also examined, as shown in Figs. 2-3. With increasing temperature, the peak intensity at either the Fe or the Co $L_{\text{III}}$ edge is becoming vanishingly small and independent of temperature above the transition temperature $T_{\text{N}}$ = 337(3) K (Fig. 3) \cite{supp}. 


Both the energy and temperature dependent studies in Figs. 2-3 demonstrate the magnetic nature of the (0, 0, 0.5) peak, which is in agreement with the AFM behavior from the magnetization in Fig. 1(c). This is further supported by studying the magnetic field $H$ dependence of the $\textbf{Q}_0$ peak on resonance (at both Fe and Co $L_{\text{III}}$ edges), examined at select temperatures (Fig. 4). The field was applied in the $ab$ plane of the sample (Fig. 1(a)). With increasing magnetic field, the $\textbf{Q}_0$ peak intensity is gradually suppressed and reaches almost zero when $H$ $\ge$ $H^{\ast}$. This can be easily understood since the applied magnetic fields are polarizing the spins along the field direction. When the magnetic moments are fully polarized under $H$ $\ge$ $H^{\ast}$, the magnetic structure enters into the polarized FM state and hence the peak intensity at $\textbf{Q}_0$ disappears. By assuming that the measured peak intensity is proportional to the $c$ component of the magnetic moment squared $m_{\text{c}}^2$, which equals $M_{\text{sat}}^2 - M_{\text{ab}}^2$ from the magnetization data in Fig. 1(d), an excellent agreement is found between the field dependence of the resonant peak intensity at Fe $L_{\text{III}}$ edge and the measured magnetization squared (green line in Fig. 4(b)). Specifically, the $H^{\ast}$ values at different temperatures inferred from both the RMXS data and the magnetization data also match each other excellently, which are summarized and plotted as the phase diagram in Fig. 4(c).

Combining all evidence, our study confirms the AFM nature of Co45-F5GT with a propagation vector $\textbf{Q}_0$ = (0, 0, 0.5) and a transition temperature $T_{\text{N}}$ $\sim$ 340 K. The similar behavior of the magnetic peak at both the Fe or Co $L$ edges indicates an important contribution from the Co spin moments in Co-F5GT. This explains why the saturation moment of Co45-F5GT at 10 K (Fig. 1(d))--$M_{\text{sat}}\sim$ 8 $\mu_{\text{B}}$ per formula unit ($f.u.$)--is only slightly smaller than the value of $M_{\text{sat}}\sim$ 10 $\mu_{\text{B}}$ per $f.u.$ in F5GT ($x$ = 0). By assuming the average Fe magnetic moment is unchanged ($\sim$ 2 $\mu_{\text{B}}$/Fe) \cite{May_2019_ACSnano_F5GT, Zhang_2020_PRB_F5GT, Zhang_2021_FCGT, Chen_2022_PRL}, the average Co spin moment contribution to the magnetization is estimated to be $\sim$ 1.1 $\mu_{\text{B}}$/Co in Co45-F5GT. Clearly, identifying the presence of an ordered magnetic moment at the Co site is an important result of our work.

\begin{figure}[t]
\centering
\includegraphics[width= 5 cm]{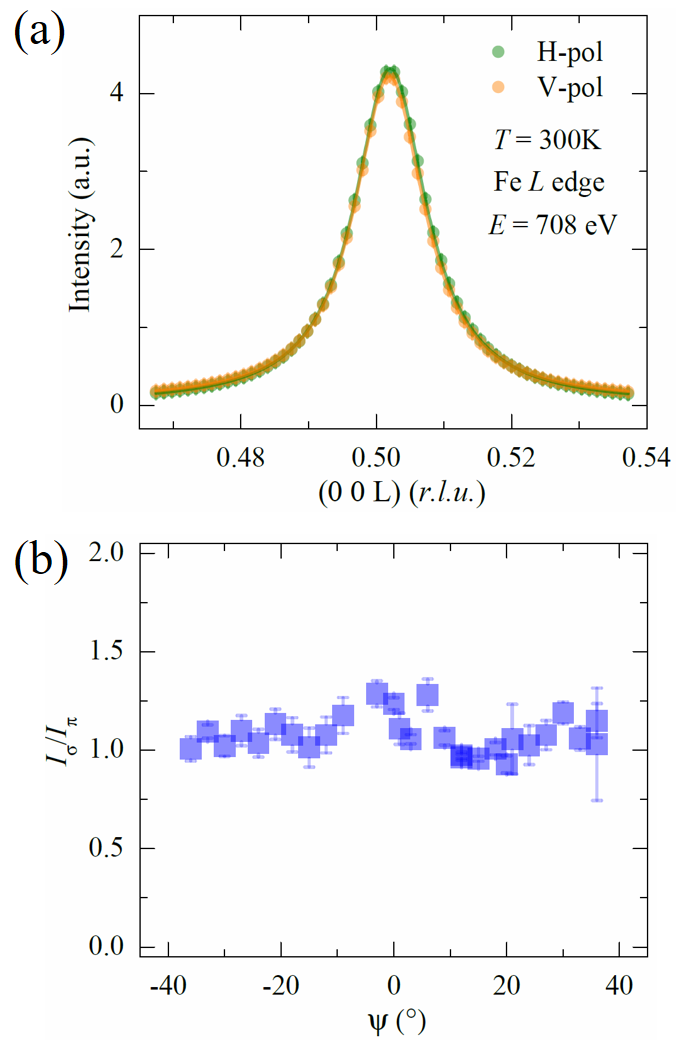}
\caption{(Color online) (a) The (0, 0, 0.5) peak, collected at 708 eV and 300K, with the incoming x-ray either horizontally ($\pi$-pol) or vertically ($\sigma$-pol) polarized. $I_{\sigma}$ = $I_{\pi}$ is observed. (b) The intensity ratio $I_{\sigma}$/$I_{\pi}$ $\approx$ 1 in a wide range of azimuthal angle $\Psi$, where $\Psi$ = 0 is defined when the [1 0 0] direction is parallel to the scattering plane.}
\label{fig:Fig5_alpha}
\end{figure}

Another important piece of information is the magnetic moment direction, specifically the magnetic structure below $T_{\text{N}}$. This can be inferred from the representational analysis \cite{Wills2000}, the photon polarization and azimuthal dependence of the RMXS data. The magnetic representation of the crystallographic sites of Fe or Co can be decomposed in terms of the irreducible representations (IRs) \cite{Wills2000}, with the propagation vector $\textbf{Q}_0$ = (0, 0, 0.5) (Table I): $\Gamma_{\text{Mag}} = 1\Gamma_{2}^{1}+1\Gamma_{3}^{1} + 1\Gamma_{5}^{2}+1\Gamma_{6}^{2}$, where $\Gamma_{2}$ , $\Gamma_{3}$ are one-dimensional IRs with moments pointing parallel to the $c$ axis and $\Gamma_{5}$ , $\Gamma_{6}$ are two-dimensional IRs with basis vectors lying in the $ab$-plane. 

Considering the AFM nature, only the solutions from $\Gamma_{3}$ and $\Gamma_{5}$ are possible. Experimentally, the magnetic peak intensity at the $\textbf{Q}_0$ position remains the same with either $\sigma$ or $\pi$ polarized incoming x-rays  (Fig. 5(a)), which implies $I_{\pi}$ = $I_{\sigma}$:

\begin{equation}
 |  \sum_{j} e^{ i \textbf{Q} \cdot \textbf{r}_j}  \textbf{k}_f \cdot \textbf{m}_j  |^2 
 = |  \sum_{j} e^{ i \textbf{Q} \cdot \textbf{r}_j}  \textbf{k}_i \cdot \textbf{m}_j  |^2 + |  \sum_{j} e^{ i \textbf{Q} \cdot \textbf{r}_j}  \textbf{e}_{\sigma} \cdot \textbf{m}_j  |^2
\label{eq:eqn2}
\end{equation}

where $\textbf{k}_i$ and $\textbf{k}_f$ are the unit vectors along the incoming and reflected photon wave-vector directions, respectively. The azimuthal $\Psi$ dependence of the $\textbf{Q}_0$ peak was performed by rotating the sample about the $\textbf{Q}_0$ direction, which is parallel to the $c$-axis (Fig. 1(a)), and measuring the peak intensity for both polarizations. Clearly, $I_{\sigma}$ $\approx$ $I_{\pi}$ is maintained over a broad range of the $\Psi$ angle (Fig. 5(b)). This rules out the possible contribution from $\Gamma_{5}$ since the presence of in-plane moments will manifest as an angle-dependent azimuthal scan by virtue of rotating the projections shown in Equation (\ref{eq:eqn2}). It is evident that only the solution from $\Gamma_{3}$ satisfies Equation (\ref{eq:eqn2}) ($I_{\pi}$ = $I_{\sigma}$), regardless of the $\Psi$ angle, since the magnet moments pointing along the $c$ direction are parallel to the rotation axis along the $\textbf{Q}_0$ direction. From these arguments, the experimental data determine an A-type AFM order with Ising moments in Co45-F5GT, as depicted in Fig. 1(b).

\begin{table}[t]
\begin{tabular}{c|cc|ccc}
\hline
  IR  &  BV  &  Atom & \multicolumn{3}{c}{BV components}\\
      &      &             &$m_{\|a}$ & $m_{\|b}$ & $m_{\|c}$   \\
\hline
$\Gamma_{2}$ & $\bfpsi_{1}$ &      1 &      0 &      0 &      1    \\
             &              &      2 &      0 &      0 &      1    \\
\hline
$\Gamma_{3}$ & $\bfpsi_{2}$ &      1 &      0 &      0 &      1    \\
             &              &      2 &      0 &      0 &     -1    \\
\hline
$\Gamma_{5}$ & $\bfpsi_{3}$ &      1 &      0 &   -1 &      0    \\
             &              &      2 &      0 &    1 &      0   \\
             & $\bfpsi_{4}$ &      1 &     -2 &   -1 &      0    \\
             &              &      2 &      2 &    1 &      0    \\
\hline
$\Gamma_{6}$ & $\bfpsi_{5}$ &      1 &      2 &    1 &      0    \\
             &              &      2 &      2 &    1 &      0    \\
             & $\bfpsi_{6}$ &      1 &      0 &    -1 &      0    \\
             &              &      2 &      0 &    -1 &      0    \\
\hline
\end{tabular}
\caption{Basis vectors for the space group P$\bar{3}$m1 with the propagation vector $\textbf{Q}_0$ = (0, 0, 0.5). The decomposition of the magnetic representation for the Fe$1$/Co$1$ site is 
$\Gamma_{\text{Mag}} = 1\Gamma_{2}^{1}+1\Gamma_{3}^{1} + 1\Gamma_{5}^{2}+1\Gamma_{6}^{2}$. The two different atoms of the Fe1/Co1 site within the unit cell are defined: atom1, $( 0,~ 0,~ 0.2318)$; atom2, $( 0,~ 0,~ 0.7682)$. The results from the irreducible representations analysis for the other Fe/Co sites are similar.}
\label{basis_vector_table_1}
\end{table}

Our RMXS study together with the magnetization data on the Co45-F5GT single crystals confirm the long-range AFM ground state. Under moderate magnetic fields ($H < H^{\ast}$), this AFM state is still maintained, since the (0, 0, 0.5) magnetic peak is only gradually weakened in magnitude, but without becoming incommensurate or significantly broadened in peak width. This suggests that it is unlikely to have the magnetic field induced exotic spin textures in Co45-F5GT with the AA-stacking order, unlike the novel AA$'$-stacked structure hosting N$\acute{\text{e}}$el-type skyrmion lattices in Co-F5GT at $x$ = 0.50 \cite{Zhang_2021_FCGT, Zhang_2022_FCGT_sciadv}. The contrasting magnetic textures in Co-F5GT, albeit with the similar Co doping level $x$ = 0.45 or $x$ = 0.5, highlight the essential role of the underlying lattice symmetry, as well as the contribution from the Co spin moments. Additionally, the Co magnetic moments appear to play an essential role for the enhancement of the magnetic transition temperature in Co-F5GT, in contrast to the nonmagnetic dilution in Ni doped F5GT \cite{Chen_2022_PRL}.

In summary, our RMXS study together with the magnetization measurements on the Co45-F5GT sample confirms the Ising nature of the A-type AFM spin structure, with a propagation vector $\textbf{Q}_0$ = (0, 0, 0.5) and a N$\acute{\text{e}}$el temperature $T_{\text{N}}$ $\sim$ 340 K. The unique, element specific characteristics of RMXS, observed through tuning the photon energy, highlight the sizable contribution from the Co spin moments to the AFM order. In addition, the magnetic ground state under the in-plane magnetic fields has also been investigated, suggesting the critical role of the underlying lattice symmetry for stabilizing unusual spin textures. Our work highlights the applicability of the RMXS technique to study the magnetic properties in Co-F5GT and other quasi-2D vdW magnets.

\bigbreak

X.C. wishes to thank Fanny M. Rodolakis for assistance before and during the RMXS experiment carried out at the Advanced Photon Source. Work at University of California, Berkeley and the Lawrence Berkeley National Laboratory was funded by the U.S. Department of Energy, Office of Science, Office of Basic Energy Sciences, Materials Sciences and Engineering Division under Contract No. DE-AC02-05-CH11231 within the Quantum Materials Program (KC2202). Work at UC San Diego was supported by the National Science Foundation under Grant No. DMR-2145080. This research used resources of the Advanced Photon Source, a U.S. Department of Energy (DOE) Office of Science User Facility at Argonne National Laboratory and is based on research supported by the U.S. DOE Office of Science-Basic Energy Sciences, under Contract No. DE-AC02-06CH11357. We thank HZB for the allocation of synchrotron radiation beamtime. A.F. was supported by the Alfred P. Sloan Foundation (FG-2020-13773) and the Research Corporation for Science Advancement via the Cottrell Scholar Award (27551).

\bibliography{FeCo50J_main.bib}

\begin{thebibliography}{51}
\bibitem [{sup()}]{supp}%
  \BibitemOpen
  \href@noop {} {}\bibinfo {note} {The $\sim$ 4 K smaller $T_{\text{N}}$, when
  comparing the number from the scattering data to the magnetization study, is
  likely due to the errors from the different thermal couples and/or the beam
  damage to the samples.}\BibitemShut {Stop}%
\end{thebibliography}

\end{document}